\documentstyle[11pt,newpasp,twoside,epsf]{article}
\def\edcomment#1{\iffalse\marginpar{\raggedright\sl#1\/}\else\relax\fi}
\reversemarginpar

\begin{document}

\title{$\beta$ Cephei and SPB stars in the Large Magellanic Cloud}
 \author{Z.\,Ko{\l }aczkowski, A.\,Pigulski}

\affil{Wroc{\l }aw University Observatory, Kopernika 11,
51-622 Wroc{\l }aw, Poland}

\author{I.\,Soszy\'nski,
A.\,Udalski, M.\,Szyma\'nski, M.\,Kubiak, K.\,\.Zebru\'n,
G.\,Piet\-rzy\'n\-ski, P.R.\,Wo\'zniak, O.\,Szewczyk,
{\L }.\,Wyrzykowski\\ (the OGLE team)}

\affil{Warsaw University Observatory, Al.\,Ujazdowskie 4, 00-478
Warsaw, Poland}

\begin{abstract}
This is a progress report of the study of pulsating main-sequence
stars in the LMC.  Using the OGLE-II photometry supplemented by the
MACHO photometry, we find 64 $\beta$ Cephei stars in the LMC.  Their
periods are generally much longer than observed in stars of this type
in the Galaxy (the median value is 0.27~d compared with 0.17~d in the
Galaxy).  In 20 stars with short periods attributable to the
$\beta$~Cephei-type instability, we also find modes with periods
longer than $\sim$0.4 d.  They are likely low-order $g$ modes, which
means that in these stars both kinds of variability, $\beta$~Cephei
and SPB, are observed.  We also show examples of the multiperiodic SPB
stars in the LMC, the first beyond our Galaxy.
\end{abstract}

\section{Metallicity and pulsations of stars in the upper part of the
main sequence}

Presently we know of about 90 $\beta$ Cephei and 100 SPB stars in the
Galaxy.  Their pulsational instability is caused by the $\kappa$
mechanism working at temperature of about 2 $\times$ 10$^5$ K (see,
e.g., Dziembowski \& Pamyatnykh 1993).  Since the opacity bump driving
pulsations in these stars originates as a result of a large number of
bound-bound transitions in the iron-group ions, the metallicity
dependence of the instability strips is an obvious consequence.
Pamyatnykh (1999) showed this in detail, finding that the
$\beta$~Cephei instability strip practically vanishes at $Z<$ 0.01.  A similar
shrinking of the instability strip is predicted for SPB stars,
although the dependence is not so strong and even at $Z$ = 0.01 the
instability strip is still quite large.

Observationally, this dependence was confirmed by Pigulski et
al.~(2002) who found a striking difference between the incidence of
$\beta$~Cephei stars in northern and southern open clusters.  This
fact could be explained as a result of the metallicity gradient in the
Galaxy.  It is therefore very important to know whether, and how many
$\beta$~Cephei and SPB stars could be found in objects of even lower
metallicities.  With their low metallicities, the Large (LMC, $Z_{\rm
LMC} \simeq$ 0.008) and Small (SMC, $Z_{\rm SMC} \simeq$ 0.004)
Magellanic Clouds are among the best objects for such a study.

\section{Data selection and analysis}

The Magellanic Clouds were the targets of many observational surveys,
including the microlensing ones, OGLE and MACHO, that provided
photometric data for millions of stars.  Recently, the OGLE-II
observations of Magellanic Clouds have been reprocessed and the
photometry of all stars in the reference frames has been obtained by
means of the modified DIA software of Wo\'zniak (2000).  We used this
photometry to search for periodic variations of stars located in the
upper main-sequence of the LMC.  For over 75\,000 stars with $V<$ 18
and $(V-I)<$ 0.5 we calculated Fourier periodograms in the range 0 to
20~d$^{-1}$.  Light curves of stars showing dominant periods shorter
than 0.35~d were checked in detail.

\begin{figure}
\plottwo{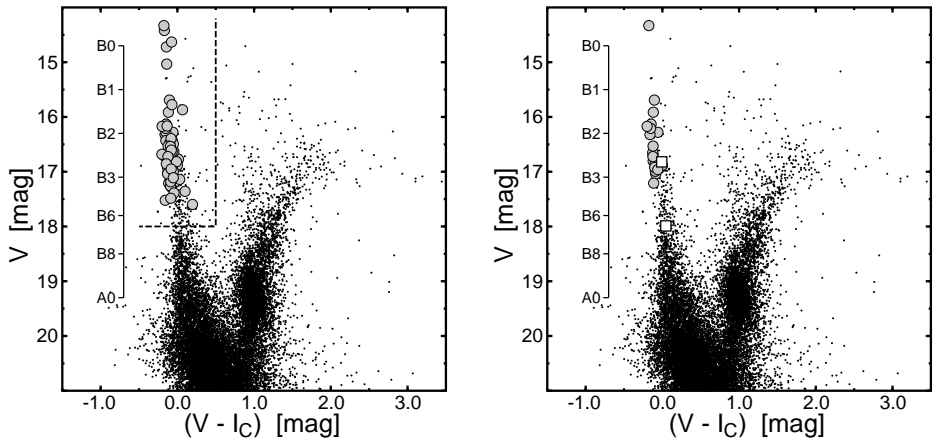}{kol_f1b.ps}
\caption{{\it Left:} Colour-magnitude diagram for the LMC showing the
positions of the 64 $\beta$~Cephei stars (grey dots) we found.  The
dashed line shows the limits of magnitudes and colours of the analyzed
stars.  Main-sequence spectral types are also given as a guide.  {\it
Right:} The same as in the left diagram, but only stars with long
periods are plotted as grey dots.  Two examples of SPB stars mentioned
in the text are shown as open squares.}

\end{figure}

\section{ $\beta$ Cephei stars in the Large Magellanic Cloud}

The first three $\beta$~Cephei stars in the LMC were found by Pigulski
\& Ko{\l }aczkowski (2002) in the OGLE-II catalogue of variable
candidates (\.Zebru\'n et al.~2000).  The new analysis of the OGLE-II
data led us to the discovery of 64 $\beta$ Cephei stars including the
three we found earlier.  We also used the MACHO data (Allsman \&
Axelrod 2001).  Detected frequencies were confirmed in all stars with
the available MACHO photometry ($\sim$80\% of the sample).  In the
final analysis we combined data from both surveys.

In the colour-magnitude diagram of the LMC (Fig.~1) $\beta$~Cephei
stars are located in the range 14 $<V<$ 17.5.  The most interesting
discovery, however, are the periods detected in these stars (Fig.~2).
In 31 out of 64 stars, we detected only a single mode.  The remaining
stars are multiperiodic with up to five modes detected.  An example of
a multiperiodic $\beta$~Cephei star is shown in the amplitude spectra
in Fig.~3.

Fig.~2 shows several interesting features:  (i) Short periods in the
LMC $\beta$~Cephei stars are generally longer than in the Galactic
ones.  The median value is 0.17~d for the Galaxy and 0.27~d for the
LMC.  (ii) There is a correlation between periods and magnitudes among
stars with 16.0 $< V <$ 17.5.  We expect this type of behaviour,
because for a given mode the period generally lenghtens with
increasing radius.  (iii) In 20 stars we detected periods which are
1.7 -- 2.2 times longer than the short period(s).  They are shown in
Fig.~2 as open squares and seem to follow some period-magnitude
relation too.  Such long periods have not been observed in the
Galactic $\beta$~Cephei stars.  They can probably be explained in
terms of the low-order $g$ modes, so the stars can be regarded as
both $\beta$~Cephei and SPB stars.

\begin{figure}
\plotone{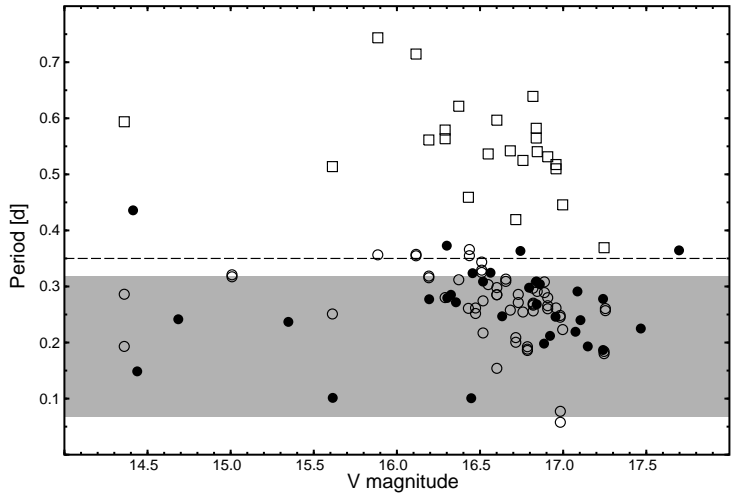}
\caption{Periods detected in 64 $\beta$ Cephei stars plotted against
the $V$ magnitude.  Periods in monoperiodic stars are shown as filled
circles, those found in multiperiodic stars as open circles (short
periods) and open squares (long periods).  The grey strip shows the
range of periods observed in the Galactic $\beta$~Cephei stars.}
\end{figure}

\section{The first extragalactic SPB stars}

In addition to the above-mentioned $\beta$~Cephei stars showing
SPB-type behaviour, we found some `classical' SPB stars in the LMC,
that is multiperiodic stars showing only periods longer than
$\sim$0.5~d.  An example of an early B-type SPB star is shown in
Fig.~3, but we also found such stars among late B-type stars (see
Fig.~1).  The complete search for SPB stars in the LMC is not yet
finished because it requires the analysis of thousands of fainter
stars and also bright stars with periods longer than 0.35~d.  This
work is in progress now and will be published separately.

\begin{figure}
\plottwo{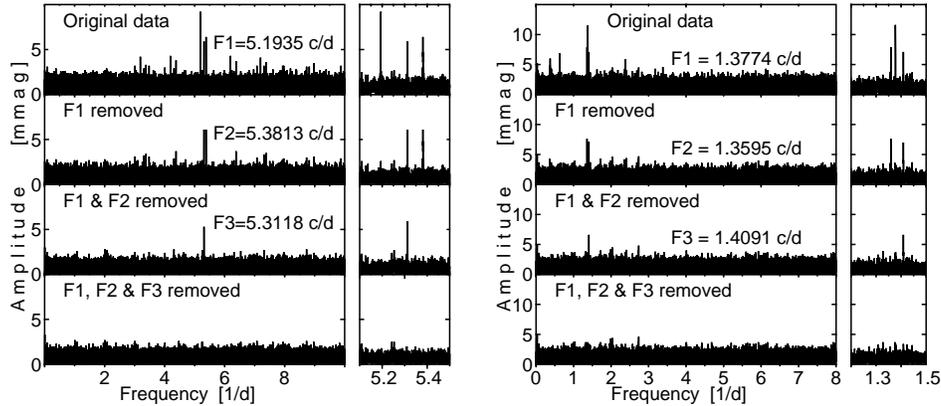}{kol_f3b.ps}
\caption{Fourier periodograms of the multiperiodic $\beta$~Cephei star
SC1-108285 (left) and SPB star SC1-223227 (right) in the LMC.  The
consecutive steps of prewhitening are shown from top to bottom.}
\end{figure}

\section{Conclusions}
Despite the large number of variable stars we found, the incidence of
$\beta$~Cephei stars in the LMC is clearly lower than in the Galaxy.
This confirms the strong dependence of the driving mechanism on
metallicity.  A detailed modelling of the observed periods, especially
the long periods found in $\beta$~Cephei stars, will probably allow
constraints to be put on the metallicities and evolutionary status of
these variables in the LMC.  Furthermore, we will extend our analysis
both to longer periods and fainter stars.  This will surely lead to
the discovery of a large number of SPB stars, but will also increase
the number of stars showing simultaneously $\beta$~Cephei and SPB-type
pulsations.

{\bf Acknowledgement.} This work was supported by the KBN grant
5\,P03D 014\,20.

\end{document}